\documentclass{IEEEcsmag}

\usepackage[colorlinks,urlcolor=blue,linkcolor=blue,citecolor=blue]{hyperref}
\expandafter\def\expandafter\UrlBreaks\expandafter{\UrlBreaks\do\/\do\*\do\-\do\~\do\'\do\"\do\-}
\usepackage{upmath,color}
\usepackage[hang,flushmargin]{footmisc}

\usepackage[super]{natbib}

\jvol{XX}
\jnum{XX}
\paper{8}
\jname{Publication Name}

\begin{document}

\sptitle{Accepted to IEEE Internet Computing, 2025}

\title{When Generative AI Meets Extended Reality: Enabling Scalable and Natural Interactions}

\author{Mingyu Zhu, Jiangong Chen, and Bin Li}
\affil{Pennsylvania State University, University Park, PA, 16801, USA}

\markboth{PREPRINT}{PREPRINT}

\begin{abstract}
\looseness-1 Extended Reality (XR), including virtual, augmented, and mixed reality, provides immersive and interactive experiences across diverse applications, from VR-based education to AR-based assistance and MR-based training. However, widespread XR adoption remains limited due to two key challenges: 1) the high cost and complexity of authoring 3D content, especially for large-scale environments or complex interactions; and 2) the steep learning curve associated with non-intuitive interaction methods like handheld controllers or scripted gestures. Generative AI (GenAI) presents a promising solution by enabling intuitive, language-driven interaction and automating content generation. Leveraging vision-language models and diffusion-based generation, GenAI can interpret ambiguous instructions, understand physical scenes, and generate or manipulate 3D content, significantly lowering barriers to XR adoption. This paper explores the integration of XR and GenAI through three concrete use cases, showing how they address key obstacles in scalability and natural interaction, and identifying technical challenges that must be resolved to enable broader adoption.
\end{abstract}

\maketitle
\chapteri{E}xtended Reality (XR) technologies, including Virtual Reality (VR), Augmented Reality (AR), Mixed Reality (MR), and everything in between, are rapidly transforming how users interact with digital content by delivering immersive, interactive experiences. XR applications span a wide range of domains, including VR-based education, AR-based assistance, and MR-based training. These immersive systems offer the potential to enhance understanding, improve task performance, and unlock new forms of communication and collaboration.

However, the widespread adoption of XR by everyday users remains limited due to several key obstacles. One of the most significant is the cost and complexity of authoring 3D content. Crafting virtual scenes, filling them with realistic objects, and enabling dynamic behaviors typically require domain expertise in 3D modeling, animation, and programming. These obstacles become especially significant when XR systems are expected to operate at scale, adapt to diverse environments, or support personalized interactions. In addition, many current XR platforms depend on rigid, non-intuitive interaction methods such as handheld controllers, pre-defined gestures, or scripted voice commands. These input constraints are often non-intuitive, creating a steep learning curve for new users and limiting the flexibility of real-time interactions.

Recent advances in Generative AI (GenAI) provide promising solutions to address these limitations. By leveraging powerful Machine Learning (ML)-based models, such as vision-language models for understanding and diffusion models for high-fidelity 3D content generation, GenAI enables users to create immersive content and interact with XR systems through more natural inputs like free-form voice and hand gestures. For example, a user could describe a desired 3D scene or ask for a virtual guide, and the system could dynamically generate or adjust content without requiring manual configuration. GenAI also offers the potential to interpret user intent, provide real-time multimodal feedback, and deliver personalized experiences based on context and past behavior.

In this article, we explore the integration of GenAI and XR and present a structured analysis of how GenAI can enhance XR-based education, assistance, and training. Through real-world examples with deployable systems, we examine how GenAI helps overcome the obstacles related to scalability and natural interaction, such as content authoring, limited interaction methods, and personalization challenges. We also identify critical system-level and technical challenges such as hallucination, latency, system resource contention, privacy, and trustworthiness, all of which require careful attention to ensure the seamless and effective integration of GenAI with XR systems. We share these perspectives to encourage further exploration and innovation at the intersection of GenAI and XR to accelerate the adoption of XR by everyday users.

\section{XR User Cases and Main Obstacles}
XR technologies are reshaping how users engage with digital content in real-world contexts. We highlight three XR use cases: VR-based education, AR-based assistance, and MR-based training. Each case is selected to represent a core category of XR, along with the unique obstacles encountered by current users.

\subsection{VR-based Education}
VR-based education creates immersive virtual environments to support teaching and learning across various subjects and skill levels. Students and teachers wear VR headsets to explore shared 3D content and environment, enabling experiential learning through interaction, observation, and audio communication. Applications range from virtual classrooms to interactive science and engineering laboratories. By placing learners in engaging, interactive 3D environments, VR enhances understanding of curriculum content, particularly for complex spatial concepts that are difficult to convey through traditional methods. For example, when introducing a new molecule in chemistry--often abstracted in 2D blackboard sketches, textbook images, or 3D slide renderings--a teacher in VR can guide students to explore and manipulate a fully interactive 3D model.
This immersive experience enhances conceptual understanding and engagement. It also provides a cost-effective alternative to physical models, which are often expensive, fragile, or unavailable in most classrooms.
 
\begin{figure}
\centerline{\includegraphics[width=18.5pc]{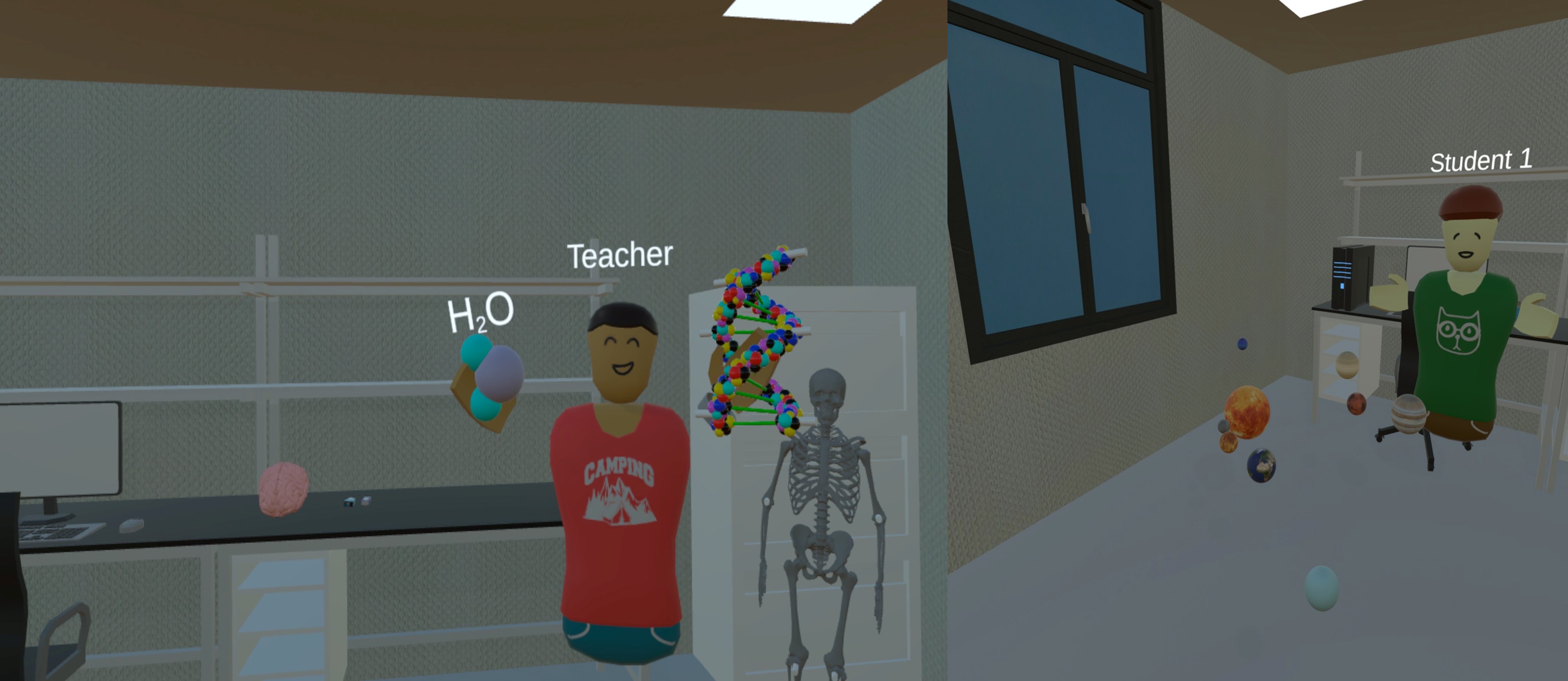}}
\caption{VR-based education. On the left, a teacher is presenting molecular and biological concepts, including a water molecule and a DNA double helix. On the right, a student is engaged in an interactive solar system model, observing planets along their animated orbits. (Some models are downloaded from Udemy.\cite{udemy})}\vspace*{-5pt}
\label{fig:vr}
\end{figure}

Figure~\ref{fig:vr} illustrates a prototype system\cite{chen2021} we developed to demonstrate the practical application of VR-based education. It shows a scenario where multiple users wear VR headsets to enter a shared virtual lab that maps the physical lab layout, including desks and chairs. Users select a teacher or student role through the interface using a controller, and each user appears as a distinct avatar, enabling visual differentiation and real-time collaboration. Within the virtual lab, students can interact with 3D models of complex scientific concepts, such as molecules, DNA, and the solar system.
These models can be explored freely from all angles without risk of damage, promoting immersive learning across subjects like biology, chemistry, and astronomy. The approach also extends to engineering, with examples such as aircraft engines, power generators, and visualizations of the electromagnetic spectrum.
The system further provides 3D drawing tools to visualize abstract concepts, such as particle trajectories in an electromagnetic field, and supports audio communication for real-time interaction and remote access. Together, these features foster immersive engagement, helping learners stay focused and motivated while experiencing rich and interactive content often unavailable in traditional classrooms.

Despite these advantages, VR-based education faces several obstacles that limit broader adoption.

\begin{itemize}
\item[{\ieeeguilsinglright}] {\it High-fidelity 3D virtual scene creation: }
Creating high-fidelity 3D virtual scenes, such as classrooms, laboratories, or museum-like environments, is essential for VR-based education. These scenes need to be carefully constructed with realistic textures, proper lighting, and well-organized spatial layouts to support effective learning and interaction. Achieving the desirable level of detail and usability typically demands close collaboration among skilled 3D designers, developers, and teachers, making the process both time-consuming and resource-intensive.
    
\item[{\ieeeguilsinglright}] {\it Massive production of 3D models: } 
Scientifically accurate and visually detailed 3D educational models are important in VR-based education, as students engage with them directly to explore complex concepts. Like virtual scene creation, developing these models requires significant expertise and effort. A key obstacle is that such massive models typically need to be created in advance, making it difficult to meet different educational needs. The challenge grows when multiple variations of a concept are required to support different learning objectives. For example, if a teacher aims to illustrate both active and dormant states of a volcano, separate models must be created for each case. This need for large volumes of pre-built, customizable models poses a significant barrier to scalable and adaptable VR-based education.

\item[{\ieeeguilsinglright}] 
{\it Interactive functionality: } 
Built-in interactivity and natural input methods are critical for responsive and engaging learning. However, many 3D educational models are static by default and lack dynamic behaviors such as animations, which are essential for effective teaching. For example, when introducing the Earth-Moon relationship, a teacher may want to demonstrate that the Moon both rotates on its axis and revolves around the Earth. Enabling such interactions requires additional coding and animation work. If every model must be manually extended with basic interactive features, it becomes difficult to scale VR content across a wide range of topics. Moreover, instructional needs vary across students and teaching styles, making it impractical to pre-program all possible interactions.  To support flexible and intuitive teaching workflows, educators should be able to trigger animations or manipulate content using natural inputs such as free-form voice commands. Traditional systems typically lack those features, resulting in undesirable usability for everyday users.
\end{itemize}

\subsection{AR-based Assistance}
AR-based assistance overlays real-time, context-aware information onto the physical world. Using mobile devices, such as smartphones, AR glasses, or headsets, users can access digital guidance while staying anchored in the real world. This modality is effective for tasks such as navigation, assembly, and situational recommendations, as it augments the user's surroundings with spatially relevant digital cues. For example, when a user opens the fridge, an AR system may recognize ingredients and suggest recipes by rendering virtual arrows and descriptions, helping the user make quick and informed decisions. Another example is the prototyped AR-based LEGO assembly assistant,\cite{yao2022} which presents step-by-step virtual instructions aligned with physical pieces to guide users through the assembly process. By overlaying digital information directly onto physical environments, AR helps users perform tasks more efficiently and intuitively, reducing cognitive load and bridging the gap between digital content and real-world action. 

\begin{figure}
\centerline{\includegraphics[width=18.5pc]{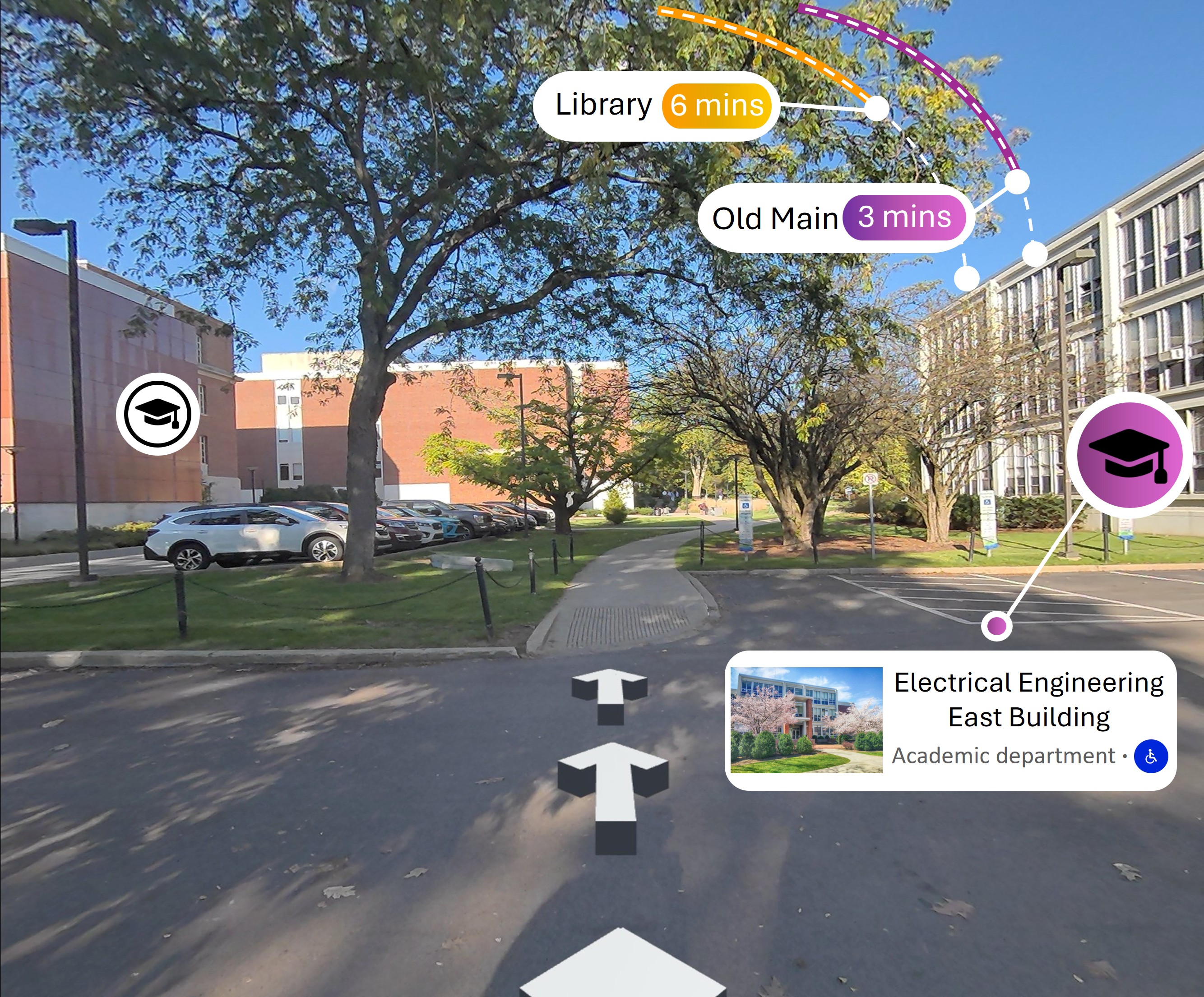}}
\caption{AR-based assistance. A student is using an AR-based navigation system on the Penn State University campus. Virtual arrows are overlaid on the road to indicate walking directions, while floating labels provide real-time guidance, including estimated time to reach the library and information about nearby buildings such as Old Main and the Electrical Engineering East Building.}\vspace*{-5pt}
\label{fig:ar}
\end{figure}

Another practical application could be a navigation system deployed at a public attraction or large venue. Using AR glasses, users can see directional arrows, labeled destinations, construction warnings, and nearby building information overlaid directly onto their surroundings. Figure~\ref{fig:ar} shows a student using such a system to find the library on the university campus. Unlike traditional navigation apps such as Google Maps, which often require users to stop and check their phones to reorient themselves, this system provides step-by-step, spatially anchored guidance that adapts in real time to the user's location and orientation. This intuitive, continuous, and context-aware visual feedback enhances wayfinding by allowing users to navigate more smoothly and confidently within complex environments. Beyond basic wayfinding, the system can offer supplementary information about nearby facilities or construction alerts, helping users better understand the environment. By embedding navigational and instructional content directly into the user’s field of view, AR-based assistance delivers a more immersive experience and reduces the mental effort required to interpret static signs, floor plans, or 2D maps.

\vspace{1em}

Although AR-based assistance offers considerable convenience, it still faces several obstacles that limit deployment and practical usability.

\begin{itemize}
\item[{\ieeeguilsinglright}] {\it Content authoring: }
In AR, authoring refers to the manual process of placing and configuring content, such as labels or 3D models, within physical environments. While current AR platforms like ARCore\footnote{https://developers.google.com/ar} and ARKit\footnote{https://developer.apple.com/augmented-reality/arkit} provide low-level capabilities such as surface detection, feature point tracking, and depth sensing to help users accurately place 3D content in the physical world, the process remains largely manual. Effective AR-based assistance systems require real-time perception and physical world understanding to automatically generate spatially relevant overlays. For example, an AR navigation system should automatically detect the floor to place directional arrows, or identify building entrances and anchor labels to doors. Relying on manual scene setup and content placement for each environment presents a major obstacle to the scalable deployment.
	
\item[{\ieeeguilsinglright}] {\it Interaction methods: }
Natural and flexible interaction is the key to effective AR-based assistance, yet many systems rely on rigid input methods, such as pre-defined triggers or handheld controllers. Those non-intuitive interaction methods could interrupt the workflow of real-world tasks and limit their usability.
Take the earlier example in which a user opens a fridge and wants a recipe suggestion. A natural interaction is simply asking, ``What can I cook with the food in the fridge''? using voice input, instead of requiring users to manually provide ingredient input by menus, buttons, or scripted commands.

\item[{\ieeeguilsinglright}] {\it Personalization and dynamic user needs: } 
A well-designed AR-based assistance system should accommodate the diverse user needs. In real-world environments, users may have various goals, expertise levels, and preferences. For example, a first-time campus visitor may need basic navigation instructions, while a returning student may prefer quick access to shortcuts or updates about temporary changes like construction zones. One-size-fits-all systems cannot adapt to these differences effectively. To provide personalized assistance, AR systems should adapt content to individuals and respond dynamically as their goals change. Without this flexibility, the system's applicability across diverse user populations is limited.

\end{itemize}

\subsection{MR-based Training}

\begin{figure}
\centerline{\includegraphics[width=18.5pc]{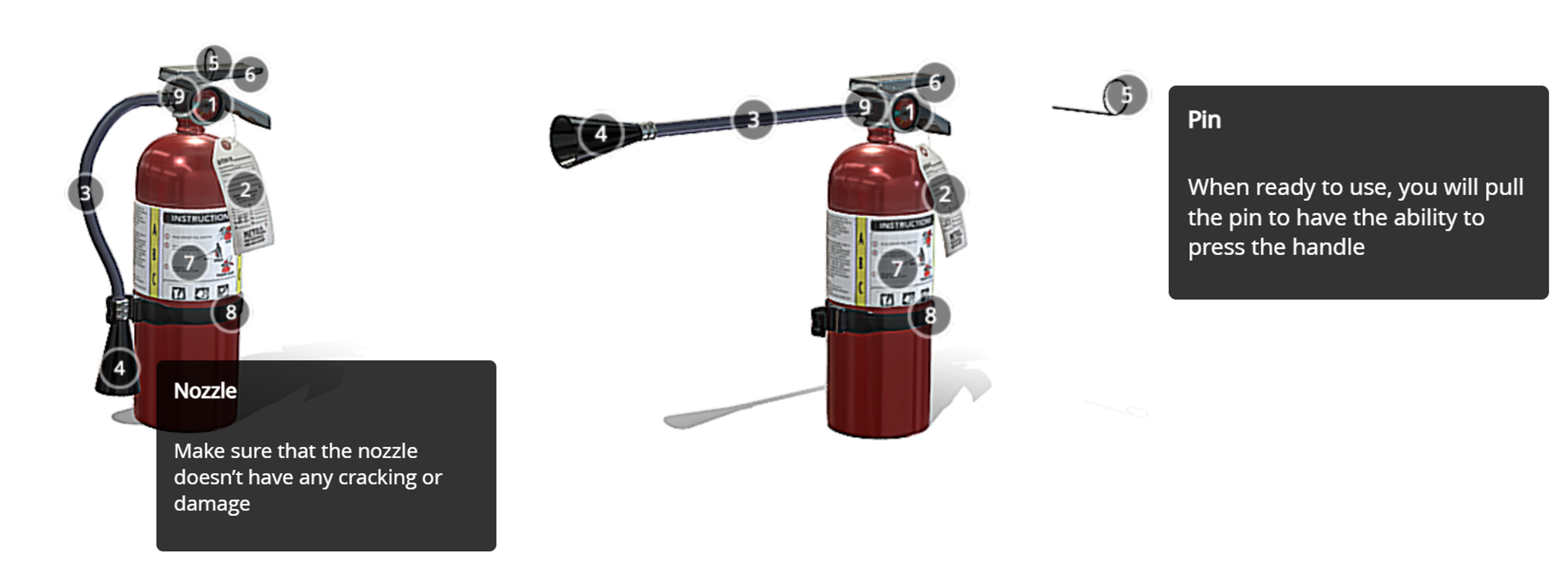}}
\caption{Virtual extinguisher model. This highly interactive virtual fire extinguisher mimics its real-world counterpart with labeled components that help users learn correct usage techniques, enabling a hands-on training experience. (This model is sourced from Sketchfab.\cite{extinguisher})}\vspace*{-5pt}
\label{fig:extinguisher}
\end{figure}

MR-based training integrates interactive virtual content into real-world settings, allowing users to engage with physical and digital elements simultaneously. Users typically rely on high-end MR devices such as the Apple Vision Pro, Meta Quest, or Microsoft HoloLens to view and interact with spatially anchored virtual objects. This hybrid interaction model supports realistic, hands-on training experiences across domains, including healthcare, manufacturing, and emergency response scenarios like firefighter training.\cite{dong2023} Unlike VR, which fully immerses users in a simulated environment, MR keeps users aware of their physical surroundings. Unlike traditional AR, where virtual objects are often limited to visual overlays without interactivity, MR enables users to manipulate virtual objects in ways that mimic interactions with their real-world counterparts. A virtual extinguisher example is shown in Figure~\ref{fig:extinguisher}. These capabilities allow users to practice complex skills in real-world scenarios, enhancing realism and improving skill transfer to actual environments, which is often difficult to achieve through purely virtual simulations or in high-risk settings.

\begin{figure}
\centerline{\includegraphics[width=18.5pc]{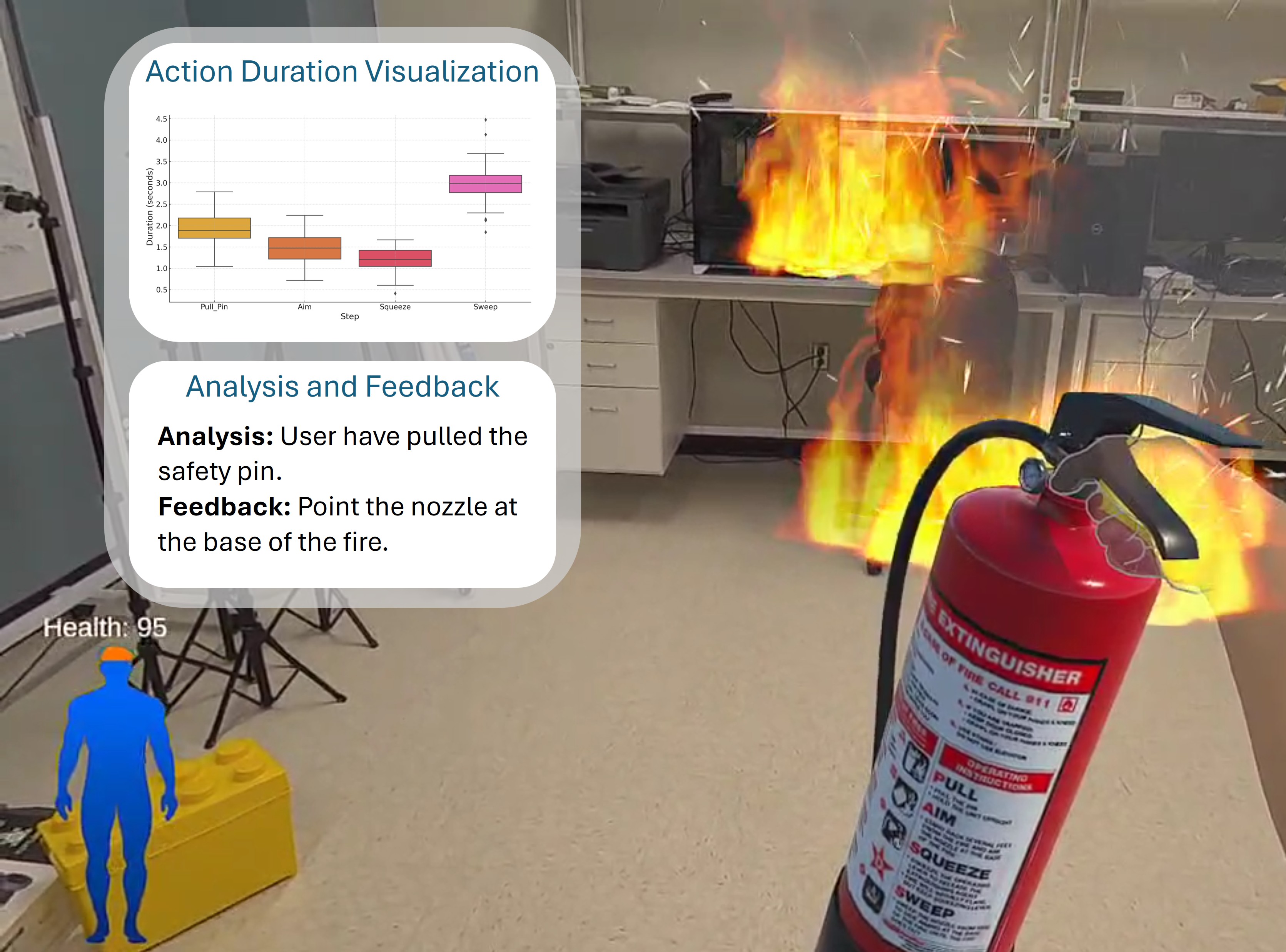}}
\caption{MR-based training. The user interacts with a virtual fire extinguisher to put out a virtual fire in a realistic physical environment. A health indicator in the lower-left corner tracks the user’s virtual health. 3D overlays provide real-time feedback, such as detecting the pulled safety pin and guiding nozzle placement. Post-training evaluation visualizes action durations to help users review and improve performance.}\vspace*{-5pt}
\label{fig:mr}
\end{figure}

Figure~\ref{fig:mr} illustrates a collaborative firefighter training scenario, where a trainee uses virtual extinguishers to put out simulated fires overlaid on real-world locations. Each trainee can view annotated instructions, animations of proper extinguisher use, and safety warnings precisely aligned with actual equipment. Unlike paper manuals or traditional digital displays, MR enables users to follow step-by-step procedures directly within their physical environment. Trainees can also see dynamic fire and smoke effects and hear crackling sounds, creating a highly immersive experience that would be unsafe or impractical to reproduce with real flames indoors. Additionally, the state of the fire and all interactive elements are synchronized across users, enabling coordinated teamwork during training. This combination of physical engagement and real-time digital guidance improves learning outcomes and supports effective skill transfer, particularly in high-risk environments where realism and safety must be balanced. By blending real and virtual elements, MR-based training provides a unique immersive and practical approach to developing complex, hands-on skills.

While MR-based training improves training effectiveness, it still faces several obstacles in practical application.

\begin{itemize}
\item[{\ieeeguilsinglright}] {\it Training scenario configuration: }
MR-based training relies on precise alignment between virtual content and the physical environment to support highly interactive and realistic experiences. Unlike AR, where simple overlays may suffice, MR scenarios often require customized content and context-aware interactions that respond accurately to real-world spatial layouts, equipment, and user workflows. This process involves both content authoring, such as placing virtual objects at the correct positions and scales, and real-time physical world perception for accurate spatial registration and system responsiveness. Setting up these training scenarios typically requires significant time, expertise, and manual effort, especially when adapting to new environments. This high degree of customization presents a major obstacle for scalability, as predefined training scenarios are often impractical across diverse scenes and use cases.

\item[{\ieeeguilsinglright}] {\it Real-time user behavior analysis: }
To maximize training effectiveness, MR-based systems should analyze user behavior in real time and provide immediate, context-aware feedback. This includes tracking user actions, gestures, and identifying deviations from the correct procedure. Based on this, the system can provide visual or audio feedback to correct mistakes, reinforce proper techniques, or guide users through complex tasks. Without such dynamic feedback, the training experience becomes passive and less responsive to individual learning needs. However, implementing accurate and adaptive behavior analysis is challenging, as it requires real-time interpretation and intelligent response generation. These capabilities are still underdeveloped in many current MR training platforms.

\item[{\ieeeguilsinglright}] {\it User performance evaluation: }
Robust post-session evaluation is critical for assessing skill mastery, tracking progress, and adapting future training to individual needs. To achieve this, the system should provide quantitative metrics that reflect how efficiently and consistently a user completes key procedures. For example, an MR training system could generate a performance chart after each session, displaying how well users followed guidance across multiple attempts. Such feedback not only reinforces learning but also helps identify skills or steps that need further improvement. While MR-based training enables realistic and interactive practice, many current systems still lack effective mechanisms for automatic post-session evaluation.

\end{itemize}

\section{GenAI meets XR}
\begin{figure*}
\centerline{\includegraphics[width=32pc]{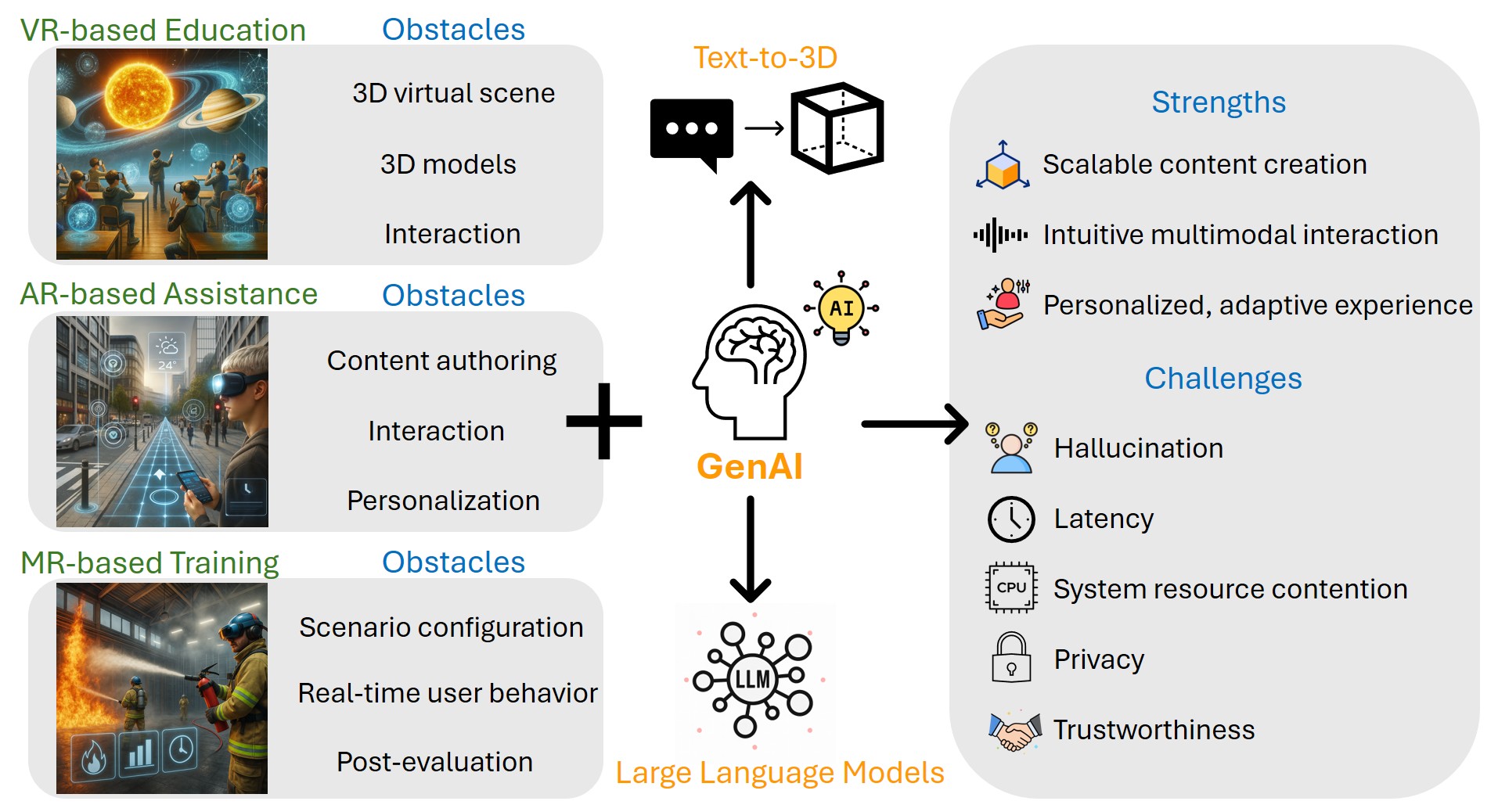}}
\caption{GenAI Meets XR. An overview of how GenAI can mitigate key obstacles in XR applications--VR-based education, AR-based assistance, and MR-based training--by enabling scalable content creation, intuitive multimodal interaction, and personalized adaptive experiences. This integration also introduces challenges such as hallucination, latency, system resource contention, privacy, and trustworthiness. (Some components of this figure were generated by OpenAI's ChatGPT.)}\vspace*{-5pt}
\label{fig:gen_xr}
\end{figure*}

While VR, AR, and MR platforms provide immersive experiences, they still face several key challenges, including labor-intensive content authoring, unnatural interaction methods, and poor scalability. GenAI introduces powerful capabilities to address these obstacles. By automatically generating 3D assets, enabling natural language interfaces, perceiving the physical world, and providing intelligent real-time feedback, GenAI greatly enhances the flexibility and usability of XR systems. Below, we discuss how GenAI helps address key obstacles in VR-based education, AR-based assistance, and MR-based training. Figure~\ref{fig:gen_xr} summarizes these obstacles, the role of GenAI, and its associated strengths and challenges across XR domains.

\subsection{GenAI-powered VR-based Education}
In VR-based education, building immersive learning environments typically requires significant time and technical skill. The learning environments should be crafted with realistic textures, lighting, and spatial layouts. GenAI can simplify this process by generating high-quality virtual scenes based on text prompts. For example, a teacher could simply request ``a virtual classroom with desks and a whiteboard,'' and the system would automatically generate a complete, interactive scene with minimal manual input. Recent text-to-3D scene generation techniques, such as RealmDreamer,\cite{shriram2025} demonstrate the ability to create detailed 3D environments from natural language descriptions without requiring scene-specific training data. Ongoing research is focused on improving scene quality and generation speed, making these techniques increasingly practical and mature.

Beyond scenes, GenAI can also support the large-scale generation of 3D educational models. Text-to-3D and Image-to-3D generation methods enable educators to quickly and cost-effectively produce diverse learning objects, especially when multiple variations of a concept are needed. For example, educators may wish to present both active and dormant volcanoes to illustrate different states. Advanced large-scale 3D synthesis systems like Hunyuan3D 2.0\cite{hunyuan2025} have demonstrated the ability to generate high-quality, textured 3D assets from simple inputs such as text descriptions or reference images, reducing the need for manual modeling.

To make VR content more engaging, GenAI can add interactive behaviors to static models. Instead of writing code, a teacher could say, ``Show how the Moon orbits the Earth,'' and the system would generate the animation. Paired with speech and gesture recognition, these capabilities allow for more natural interaction, empowering educators to control learning content without technical expertise. Recent frameworks like LLMR\cite{torre2024} and GPT-VR Nexus\cite{chen2024} demonstrate how large language models (LLMs) can generate responsive and interactive behaviors from natural language commands, such as making an object grabbable, scalable, or triggering basic animations. These interactions are achieved with low latency and error rates, making immersive engagement more reliable. 

\subsection{GenAI-powered AR-based Assistance}
In AR-based assistance, GenAI provides promising solutions to the content authoring challenge. Although platforms like ARCore and ARKit provide foundational tools for environment mapping and surface sensing, placing digital overlays such as arrows, labels, or 3D objects still relies heavily on manual configuration. GenAI can automate this process by interpreting spatial data and generating context-aware AR content. For example, instead of manually placing a label, a user could say ``Label the library entrance.'' The system could then use Vision LLMs to interpret the request, analyze the scene, and automatically generate a label on the correct door. By reducing manual effort and enabling intuitive input, GenAI makes AR-based assistance more scalable and user-friendly.

Interaction is another area where GenAI can improve the AR experience. Traditional AR systems often rely on rigid inputs such as menus, controllers, or predefined commands, which can interrupt the flow of real-world tasks and feel unnatural. GenAI enables more intuitive and flexible multimodal interactions, allowing users to communicate using natural language. For example, the LLMER system\cite{chen2025} demonstrates high interactivity by allowing users to perform tasks through natural language commands, which participants in its user study consistently rated as engaging and easy to use. Revisiting the earlier fridge example, a GenAI-powered AR assistant could visually recognize food items, understand user intent, and suggest recipes. Detailed cooking steps could then appear spatially anchored in the user’s view, while a text-to-speech model speaks the instructions like a real human, providing hands-free, multimodal guidance. By supporting natural interaction, GenAI helps make AR-based assistance more accessible, hands-free, and seamlessly integrated into everyday activities.

Another strength of GenAI is its ability to generate content in different styles and personalize the AR experience based on user needs and context. In real-world environments, users often have varying goals and preferences. GenAI can analyze user behavior, environment, and interaction history to personalize both content and presentation style. For example, it can adjust visual elements or prioritize information based on user preferences or the environment. The ViDDAR framework\cite{xiu2025} illustrates this by using vision-language models to evaluate whether virtual content is contextually appropriate and user-friendly, thereby enhancing the user experience. This user-centric approach allows AR systems to move beyond one-size-fits-all solutions and function more like intelligent, adaptive assistants.

\subsection{GenAI-powered MR-based Training}
In MR-based training, virtual objects are not just visual overlays but highly interactive, requiring precise alignment with the physical environment. Beyond spatial accuracy, effective training often demands scenario-specific customization, such as tailored guidance or safety warnings for each virtual object or training step. By interpreting real-world scenes and automating content generation, GenAI can streamline the customization of MR training environments. For example, given a 3D scan of a workspace, a GenAI-powered system could automatically align virtual objects, configure instructional overlays, and generate interactive training scenarios. The XaiR framework\cite{srinidhi2024} leverages Multimodal LLMs (MLLMs) to understand spatial environments and support accurate content authoring, with the user study showing that MLLMs were only slightly slower than human assistants. This mitigates the need for manual setup, accelerates deployment, and improves scalability across different training environments.

Another essential component of effective MR-based training is real-time user feedback. Traditional training often requires supervision or human intervention to correct mistakes and guide learners. Multimodal GenAI models, such as Vision LLMs, can interpret user actions by integrating visual, linguistic, spatial, and temporal data. This enables the system to detect procedural errors and deliver immediate, context-aware guidance. Instead of relying on human supervisors, GenAI can provide visual or audio prompts during training, offering responsive support that helps learners stay on track and reinforces correct techniques. For example, the Explainable XR framework\cite{kim2025} demonstrates how LLMs can assist in analyzing user behaviors in XR by interpreting multimodal user actions and delivering meaningful insights, which could be extended to support real-time feedback in training scenarios across diverse domains.

Post-session performance evaluation is also critical. Effective evaluation should include quantitative metrics that reflect how well users performed, such as efficiency, accuracy, and task completion time, visualized through 2D graphs or 3D charts. GenAI can support this process by automatically generating performance reports and visual progress charts. For example, after a firefighting simulation, users might receive a set of visualizations illustrating how accurately and efficiently they followed required procedures. An example is shown in Figure~\ref{fig:mr}. Automated evaluation improves visibility into skill development and helps tailor future training to individual needs, making MR-based training more adaptive and effective. Such evaluations also facilitate dynamic difficulty adjustment, enabling the system to offer more guidance for beginners or greater challenges for advanced users.

\section{Challenges and Potential Solutions}

Although GenAI brings powerful capabilities to enhance XR systems--offering promising solutions for scalability and natural interaction--several system-level and technical challenges must be addressed to achieve seamless and successful integration of GenAI with XR technologies.

\begin{itemize}

\item[{\ieeeguilsinglright}]
{\it Hallucination: }AI hallucination is an infamous issue when applying GenAI across fields. AI models sometimes generate unreliable responses, such as floating a virtual cup in the air or creating incorrect annotations of real-world objects. Those unpredictable behaviors could degrade the user experience if not carefully managed. One possible solution is explainable AI, which helps researchers better understand how these models make decisions and improve their reliability. Another approach is to pair GenAI with more focused AI systems, like object recognition models, which can act as a sanity check and help filter out faulty results.

\item[{\ieeeguilsinglright}] {\it Latency: }
Responsive XR applications require real-time feedback, especially during interactions such as issuing voice commands or manipulating virtual objects. 
Current MR systems that integrate GenAI still face substantial latency bottlenecks. For example, LLMER\cite{chen2025} reports an average speech-to-action latency of 10.35 seconds using GPT-4o, reflecting the multi-stage processing time of integrating LLMs with XR. For text-to-3D generation, even optimized pipelines such as Meshy AI\footnote{https://www.meshy.ai/blog/meshy-1-generate-3d-models-with-ai-in-just-a-minute} and Meta 3D Gen\cite{meta2024} need under one minute to produce high-fidelity 3D assets. While acceptable for design workflows, this remains far from real-time interaction. More generally, large vision–language and diffusion models deployed in the cloud exhibit latencies that vary with model size, input complexity, and network conditions, and can become substantial in practice. Such delays disrupt immersion and hinder user experience.
XR system itself also imposes strict latency constraints. Motion-to-photon latency should remain below 20 milliseconds\footnote{https://www.ericsson.com/en/reports-and-papers/ericsson-technology-review/articles/xr-and-5g-extended-reality-at-scale-with-time-critical-communication} for natural hand–object coordination, while AR-based guidance tasks such as path overlays should stay below the human reaction time of roughly 273 milliseconds.\footnote{https://humanbenchmark.com/tests/reactiontime} Meeting these real-time requirements while integrating GenAI remains a major challenge.
Addressing it requires a combination of strategies, including model optimization, edge computing, and pipeline parallelization. In practice, workloads are split across device, edge, and cloud tiers: lightweight perception on-device, mid-level fusion at the edge, and heavy generative inference in the cloud, enabling simultaneous perception, generation, and rendering.

\item[{\ieeeguilsinglright}] {\it System resource contention:}
Integrating GenAI and XR enhances immersion but also increases demands on CPU, GPU, and network resources. Even on edge servers, resource management remains critical. For example, Hunyuan3D 2.0\cite{hunyuan2025} requires 6 GB of VRAM for shape generation and 16 GB for combined shape and texture generation, showing the significant GPU memory demands of advanced models.
XR applications often run on mobile devices, which are less powerful than edge or desktop workstations, making resource management even harder. While one obvious solution is to improve hardware, techniques like intelligent task scheduling and orchestration can help optimize usage. Additionally, edge and cloud computing can lighten the load by offloading certain tasks, especially non-time-sensitive ones, from the mobile device to more powerful servers. 

\item[{\ieeeguilsinglright}] {\it Privacy: } 
Balancing GenAI's power with robust and user-centric privacy protections remains an open research challenge. GenAI-enhanced XR systems often rely on multimodal inputs such as images, voice commands, spatial layouts, and behavior. Transmitting these data to cloud-based models like ChatGPT introduces significant risks, including data leakage, unwanted exposure of personal environments, and potential extraction of private information through hidden machine learning operations. These risks are intensified in collaborative XR spaces such as classrooms or healthcare environments, where sensitive information such as facial data or patient records may be streamed for inference without explicit awareness or consent. Beyond technical challenges, these scenarios raise ethical concerns around transparency, trust, and regulatory compliance.
Recent efforts on privacy protection have explored techniques including selective masking, on-device processing, and secure inference pipelines.

\item[{\ieeeguilsinglright}] {\it Trustworthiness: } 
Trustworthiness is another growing concern in the integration of GenAI and XR. As GenAI increasingly influences what users see, hear, and interact with in immersive environments, ensuring the accuracy, reliability, and intent of generated content becomes essential. Users may be misled by hallucinated information, incorrect spatial cues, or inconsistent behaviors, leading to degraded experience or even harmful outcomes in domains like education, healthcare, or training. Moreover, the opaque nature of many generative models makes it difficult for users to understand system decisions, undermining transparency and user confidence. To foster trust, future GenAI-XR systems should incorporate mechanisms for validating outputs, explaining system behavior by visualizing intermediate results, and allowing users to verify or override generated content when necessary.

\end{itemize}

\section{CONCLUSION}
XR technologies are reshaping how people interact with the digital world. Through three representative cases--VR-based education, AR-based assistance, and MR-based training--we highlighted obstacles such as labor-intensive content creation, non-intuitive interaction, and limited personalization. GenAI offers promising solutions by enabling scalable content generation, intuitive multimodal interaction, and adaptive user experiences. However, realizing this potential requires overcoming key technical challenges, including hallucination, latency, system resource contention, privacy, and trustworthiness concerns. We hope this article provides valuable insights to inspire future research and innovation at the intersection of GenAI and XR. \vspace*{-8pt}

\section{ACKNOWLEDGMENTS}
This work has been supported in part by NSF under the grants CNS-2152658 and M3X-2420351, and DARPA grant HR0011-2420366.

\def\refname{REFERENCES}

\begin{IEEEbiography}{Mingyu Zhu}{\,} is a Ph.D. student at Pennsylvania State University, University Park, PA, 16801, USA. His research interests include extended reality and its integration with generative AI and computer vision. Zhu received his M.S. degree in Electrical and Computer Engineering from the University of Michigan. He is a student member of the IEEE and the ACM. Contact him at mintrrey@psu.edu.
\end{IEEEbiography}

\begin{IEEEbiography}{Jiangong Chen}{\,} is a Ph.D. candidate at Pennsylvania State University, University Park, PA 16801, USA. His research interests include computer networking, extended reality, and large language model agents. Chen received his M.S. degree in Electrical Engineering from the University of Rhode Island. He is a student member of the IEEE and the ACM. Contact him at jiangong@psu.edu.
\end{IEEEbiography}

\begin{IEEEbiography}{Bin Li} {\,} is an associate professor at Pennsylvania State University, University Park, PA 16801, USA. His research interests include communication networks, extended reality, digital twins, and mobile edge computing. Li received his Ph.D. degree in Electrical and Computer Engineering from The Ohio State University. He is a senior member of the IEEE and a member of the ACM. Contact him at binli@psu.edu. 
\end{IEEEbiography}

\end{document}